# Coupling Hyperspectral and 3D Data for the preventive Conservation of Palace-museums

Julie Fromager[1], Vincent Gauthier[2], Loïc Martinez[3], Danilo Forleo[5], Stéphane Serfaty[4]
[1]SATIE UMR CNRS 8029, CY Cergy Paris University, Cergy-Pontoise, 95000, FRANCE - julie.fromager@cyu.fr
[2]vincent.gauthier@cyu.fr [3]loic.martinez@cyu.fr [4]stephane.serfaty@cyu.fr
[5]Musée national des châteaux de Versailles et de Trianon, 78008 Versailles, FRANCE - danilo.forleo@chateauversailles.fr

## 1. Context: palace-museums needs in preventive conservation

Palace-museums, of which the Versailles Palace is a peculiar example, represent a particularly distinctive category of museums. In these sites, space and collections have an intrinsic relationship: artworks are displayed within their historical environment, arranged to maintain the authentic, lived-in, and heritage character of the interiors. Thus, the exhibited pieces cannot be isolated behind protective glass. Paintings, tapestries, wall hangings, ceramics, and furniture coexist within the same environment, producing a complex interplay of materials that further challenges conservation efforts. Each element presents its own conservation requirements and vulnerabilities. Moreover, lighting conditions and the constant flow of visitors introduce additional constraints. As visitors circulate among the objects and approach them closely, the potential for wear and deterioration increases (Forleo, 2020).

In light of climate change and the imperative for sustainable energy management, conventional conservation practices for monuments, historic homes, and exhibited collections must be reassessed. The new environmental context (marked by higher temperatures, increased humidity and precipitation, and abrupt, unpredictable fluctuations) no longer allows conservation and resilience strategies for historic houses to rely solely on long-term records of macroscopic environmental data (Forleo, 2020). Despite extensive preservation measures (such as climate control systems, filtered window protection, and comprehensive risk management plans), heritage residences are experiencing a growing rate of deterioration in both collections and décor. These issues appear increasingly linked to atypical climatic phenomena. Establishing the correlation between environmental variables and the condition of artworks *in situ* requires comprehensive and individualized monitoring, allowing for an understanding of cause-and-effect mechanisms. To address this challenge, the EPICO method provides a systematic framework for assessing deterioration risks in palace-museums through multi-scale monitoring and correlation between environmental parameters and object condition.

It is in this context that this research was developed as part of the ESPADON project[1], which aims to provide the community with new multi-physical imaging tools and unique digital resources in the processing and management of big data. Furthermore, it addresses the European ECHOES project objectives, for the collaborative analysis of cultural heritage asset[2].

## 2. A multi-scale gonio-hyperspectral system

This paper aims to implement novel techniques for multidimensional (spatial and spectral) *in situ* measurements and develop data processing tools to enhance existing decision support tools. Given the wide variety of studied materials, simple measurements of room temperature, humidity or sunlight exposure are not enough. It is necessary to identify the links between these environmental measurements and the condition of the objects *in situ*, but also how these conditions evolve, to hope comprehending the cause-and-effect relationships on the long term.

The scientific challenge is to establish a three-dimensional map of the physical characteristics of spaces and objects and to monitor versus time. A new hyperspectral imaging (HSI) system combined with 3D reconstruction (Figures 1, 2, 3 and 4) is presented in order to identify, in the long term, the thresholds of irreversible physico-chemical transformation of objects and the environments that constitute them. Thanks to HSI, the tools developed in this research aim to obtain detailed and objective analyses of heritage objects *in situ* and artefacts, going well beyond the conventional focus on (visible) condition reports and macroscopic temperature and humidity measurements (Huijbregts, 2015). Furthermore, the use of HSI applied to a broad range of materials is a novelty for heritage conservation (Liang, 2012; Picollo, 2020). In addition to hyperspectral analysis, the goniometric analysis considered requires precise localisation of objects in space in order to monitor them over long periods (several years).

The Versailles Palace is chosen as the most representative palace-museum, in which the Queen's Bedchamber and the Mercury Salon are the rooms studied in this research. One measurement is planned to be carried for each season, on the span of a year and a half. Three measurements campaigns have already been carried solely in the Queen's Bedchamber, during the Spring, Summer and Autumn 2025, and the next is planned in both rooms for winter 2026. For this work, the following equipment was chosen: two HS cameras (Pika XC2 and Pika IR+, covering the electromagnetic spectrum 400nm to 1700nm) and a 3D LiDAR camera (Helios2+). We chose to use two 400W halogen lamps, mounted on either side of the HS system to ensure uniform illumination of the scene in both the visible and infrared range.

## 3. Results

Typical objects and materials making up the furniture and walls of the selected rooms were scanned with the 3D HSI setup, so that any changes over time can be detected and a diagnosis can be made in correlation with the sensor data readings (temperature, hygrometry). The different types of material are subjected to different types of temporal variation or deterioration, which are recorded and monitored.

Water content in wooden materials is currently studied on an infrared level (valley point around 1450nm) both on i*n-situ* data and laboratory samples (Yuan, 2025). The study shows evidence of water inside artefacts, such as the Queen's serre-bijoux, but further investigation on longer term is needed to monitor its variations.

---

[1] Equipex+ Espadon, last visited on November 2025, https://www.sciences-patrimoine.org/espadon/

[2] Eccch. Echoes, last visited on November 2025, https://www.echoes-eccch.eu/

Furthermore, the ultimate goal of this research is to create a 3D digital twin of the studied rooms, compiling the hyperspectral data over time with their precise location and surface information. This way, a 3D model of the rooms augmented with HSI and environmental data and extracted pertinent parameters is presented. Its evolution through the seasons allows preventive conservators to have a precise overview of how the climate affected the works of art during a given period of time.

## Funding

This HypErPICO work has benefited from French State aid managed by the Agence Nationale de la Recherche under the future investment program integrated into France 2030, bearing the reference ANR 21-ESRE-0050 EquipEx+ ESPADON and École Universitaire de Recherche Paris Seine Humanités, Création, Patrimoine - Fondation des sciences du patrimoine ANR-17-EURE-0021.

## References

Forleo, D. and Francaviglia, N., ,2020 : “Conserver les collections des demeures historiques : application de la méthode d’évaluation EPICO au château de Maintenon”

Huijbregts, Z. et al, 2015: “Modelling of heat and moisture induced strain to assess the impact of present and historical indoor climate conditions on mechanical degradation of a wooden cabinet,” *J Cult Herit*, vol. 16, no. 4, pp. 419–427, doi: 10.1016/j.culher.2014.11.001.

Liang, H., 2012: “Advances in multispectral and hyperspectral imaging for archaeology and art conservation.” Appl. Phys. A 106, 309–323, https://doi.org/10.1007/s00339-011-6689-1

Picollo, M et al, 2020 : “Hyper-Spectral Imaging Technique in the Cultural Heritage Field: New Possible Scenarios.” Sensors, 20(10), 2843. https://doi.org/10.3390/s20102843

Yuan, Z. et al., 2025: “Rapid and non-destructive detection of wood density based on NIR hyperspectral imaging technology and moisture correction methods.” Spectrochim Acta A Mol Biomol Spectrosc, vol. 341, doi: 10.1016/j.saa.2025.126410.

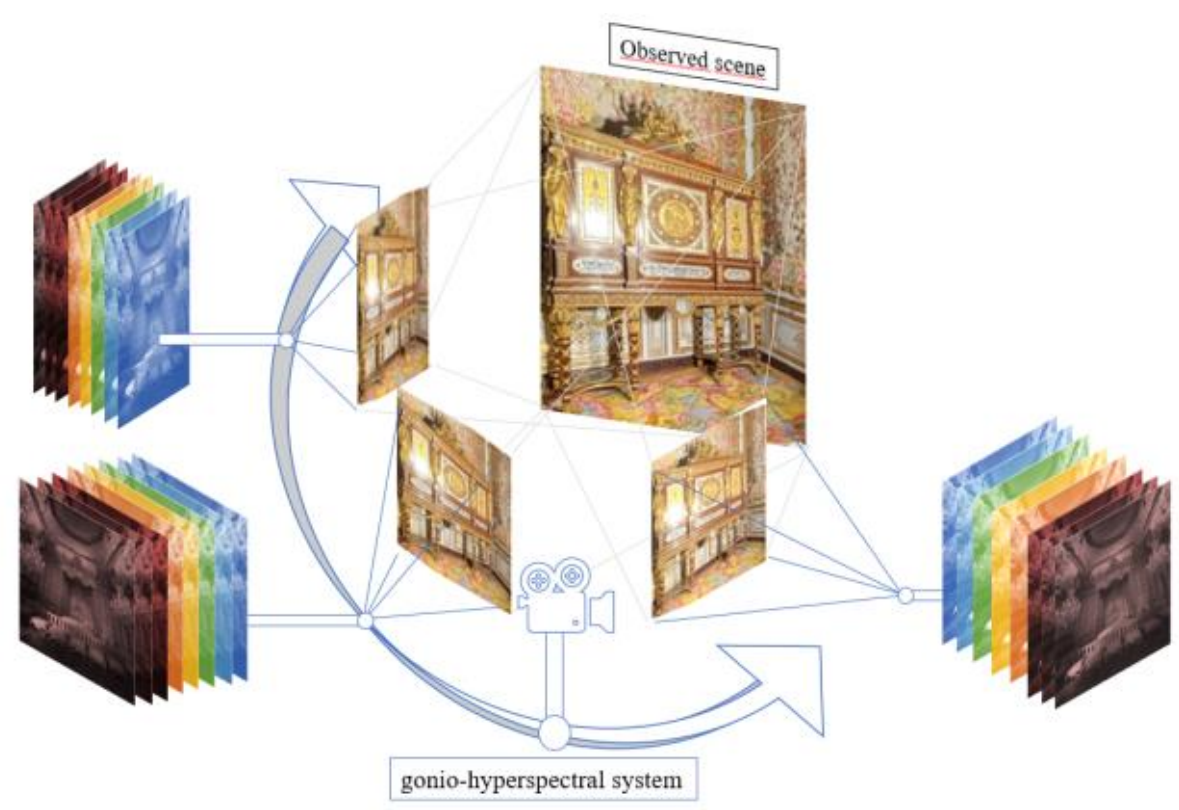

Figure 1. Gonio-Hyperspectral system

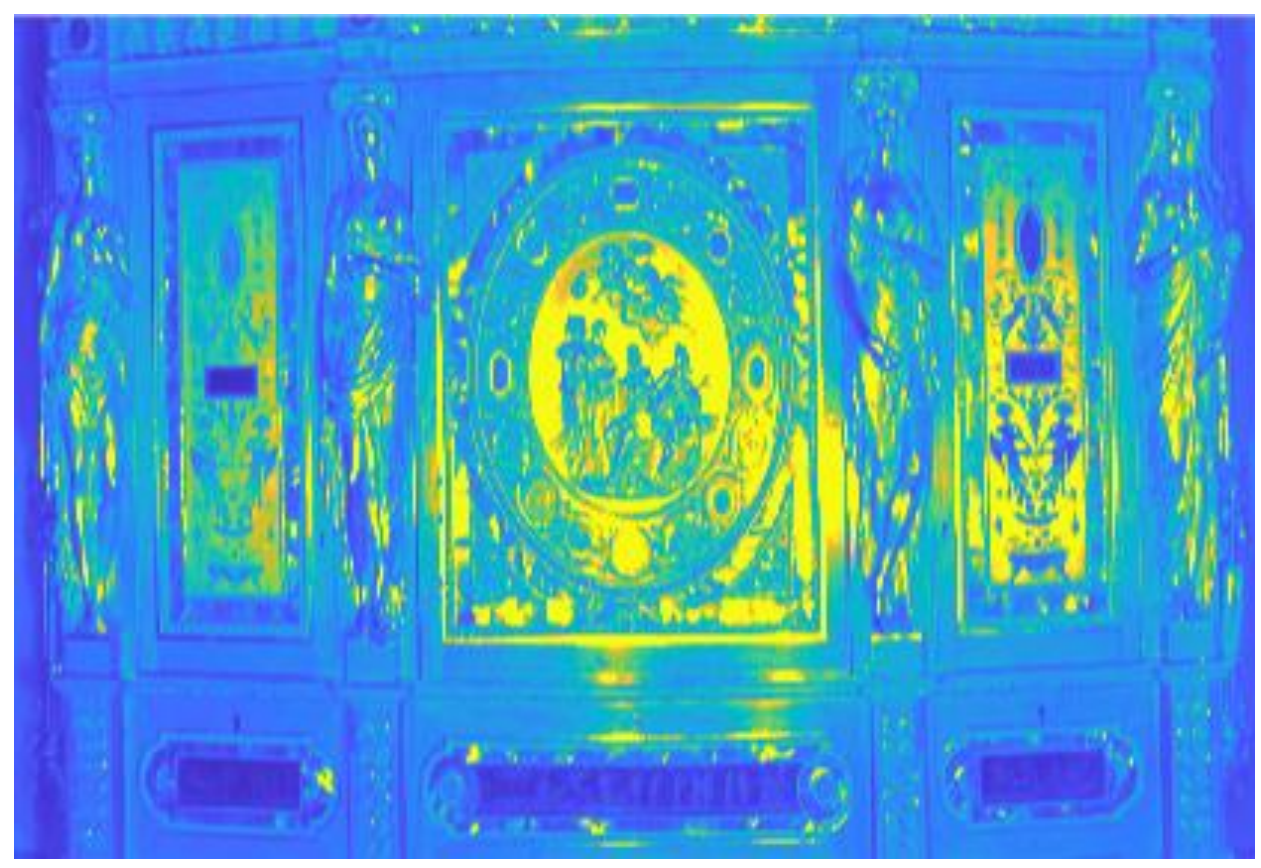

Figure 2. Hyperspectral view of the scene at 1500nm

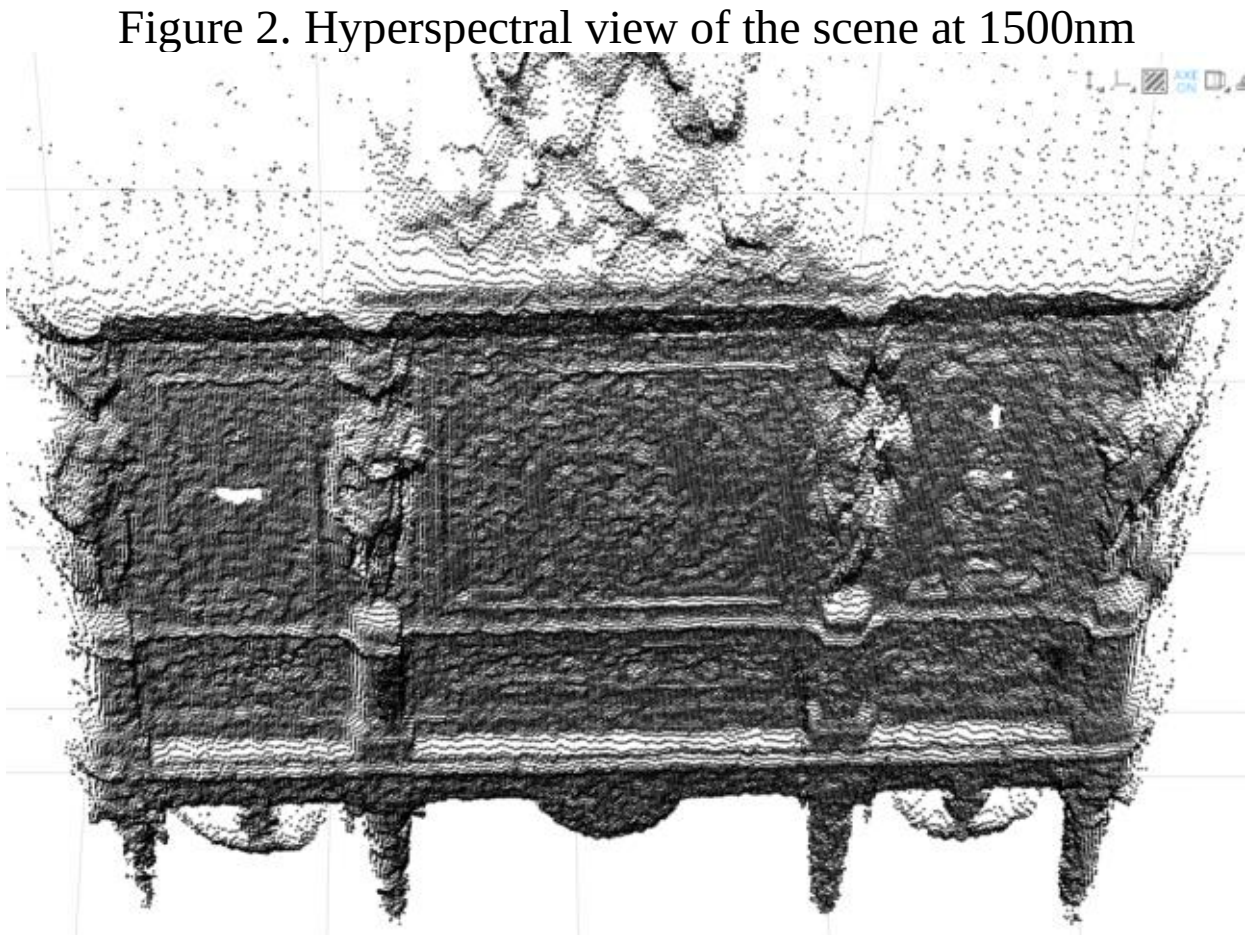

Figure 3. 3D LiDAR view of the scene

Data correlation
Statements
Water content
Discolouration
Cracks
Etc.
climate variations
Data comparison
previous data sets
new data set
Data scale alignment with previous data sets
previous data sets
new data set
Integration of new data
environmental data
climate variations
Data scale alignment
HSI VIS data
HSI IR data
3D data
Data acquisition
cameras
new data set
HSI VIS data
HSI IR data
3D data
sensors
environmental data

Figure 4. Workflow